\RequirePackage{etoolbox}\csdef{input@path}{{style/}{graphics/}}
\documentclass[aoas,MSNbibl,nameyear,dvips]{arximspdf}
\usepackage{mathbh}
\usepackage{graphicx}

\doi{10.1214/14-AOAS763} 
\volume{8}
\issue{4}
\pubyear{2014}
\firstpage{2002}
\lastpage{2026}
\docsubty{FLA}

\makeatletter
\newcommand{\indep}{\perp\!\!\!\perp}
\makeatother

\begin{document}
\begin{frontmatter}

\title{Equivalence testing for functional data with an~application to
comparing pulmonary function~devices}
\runtitle{Equivalence testing for functional data}

\begin{aug}
\author[A]{\fnms{Colin B.}~\snm{Fogarty}\corref{}\ead[label=e1]{cfogarty@wharton.upenn.edu}}
\and
\author[A]{\fnms{Dylan S.}~\snm{Small}\ead[label=e2]{dsmall@wharton.upenn.edu}}
\runauthor{C. B. Fogarty and D. S. Small}
\affiliation{University of Pennsylvania}
\address[A]{Department of Statistics\\
The Wharton School\\
University of Pennsylvania\\
Philadelphia, Pennsylvania 19104\\
USA\\
\printead{e1}\\
\phantom{E-mail:\ }\printead*{e2}} 
\end{aug}

\received{\smonth{12} \syear{2012}}
\revised{\smonth{2} \syear{2014}}

%
\begin{abstract}
Equivalence testing for scalar data has been well addressed in the
literature, however, the same cannot be said for functional data. The
resultant complexity from maintaining the functional structure of the
data, rather than using a scalar transformation to reduce
dimensionality, renders the existing literature on equivalence testing
inadequate for the desired inference. We propose a~framework for
equivalence testing for functional data within both the frequentist and
Bayesian paradigms. This framework combines extensions of scalar
methodologies with new methodology for functional data. Our frequentist
hypothesis test extends the Two One-Sided Testing (TOST) procedure for
equivalence testing to the functional regime. We conduct this TOST
procedure through the use of the nonparametric bootstrap. Our Bayesian
methodology employs a functional analysis of variance model, and uses a
flexible class of Gaussian Processes for both modeling our data and as
prior distributions. Through our analysis, we introduce a model for
heteroscedastic variances within a Gaussian Process by modeling
variance curves via Log-Gaussian Process priors. We stress the
importance of choosing prior distributions that are commensurate with
the prior state of knowledge and evidence regarding practical
equivalence. We illustrate these testing methods through data from an
ongoing method comparison study between two devices for pulmonary
function testing. In so doing, we provide not only concrete motivation
for equivalence testing for functional data, but also a blueprint for
researchers who hope to conduct similar inference.
\end{abstract}

%
\begin{keyword}
\kwd{Equivalence testing}
\kwd{functional data analysis}
\kwd{bootstrap}
\kwd{Bayesian}
\kwd{Gaussian processes}
\end{keyword}
\end{frontmatter}

\section{Introduction}\label{secintro}

An equivalence test is a statistical hypothesis test whose inferential
goal is to establish practical equivalence rather than a statistically
significant difference [\citet{ber96}]. These tests arise from the fact
that within the frequentist paradigm, failing to reject a null
hypothesis of no difference is not logically equivalent to accepting
said null. Examples of scenarios requiring equivalence tests include
the assessment of a~generic drug's performance relative to a~brand name
drug and method comparison studies, in which the agreement of a~new
device with the ``gold-standard'' for measuring a particular phenomenon
must be assured before the new device can replace the old one.

Equivalence tests for scalar data typically involve the establishment
of upper and lower equivalence thresholds dependent on the metric of
equivalence being used. The inferential aim is to establish that the
metric falls within the upper and lower equivalence thresholds with a
prespecified Type I error rate. See \citet{ber96} for a comprehensive
overview of commonly used procedures. Oftentimes the use of scalar data
is adequate, but in some instances the question of practical
equivalence cannot be reduced to a hypothesis regarding scalar data.

The motivation for this research arose from a method comparison study
between a new device for assessing pulmonary function, Structured Light
Plethysmography (SLP), and the industry standard for such assessments,
a spirometer. SLP holds many advantages over spirometry: it is
noninvasive, it can be used to diagnose patients of a wider range of
age and health levels, and it provides detailed information regarding
specific regions of the lung that may be malfunctioning. Before SLP may
be used extensively for diagnostic purposes it must be assured beyond a
reasonable doubt that the measurements obtained by SLP are practically
equivalent to those produced by a spirometer.

\begin{figure}

\includegraphics{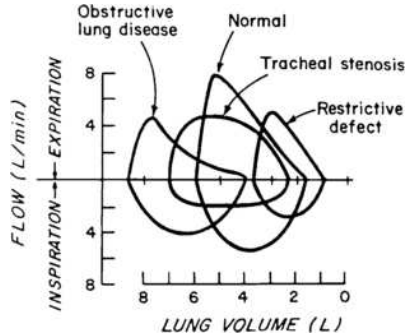}

\caption{Flow-Volume Loops corresponding to various levels of pulmonary health, calculated
using a spirometer.
[Reproduced from Goudsouzian,  N.  and Karamanian, A. (\citeyear{loop}).
\textit{Physiology for the Anesthesiologist}, 2nd ed.  Appleton-Century-Crofts, Norwalk, CT,
with permission.]}
\label{figdisease}
\end{figure}

Doctors rely on a host of information that can be produced both by SLP
and by spirometry. Some of these measurements are scalar and, hence,
their equivalence can be addressed using available scalar methods;
however, not all diagnostic tools utilized are scalar. For example, the
``Flow-Volume Loop'' is a phase plot of flow of air into and out of the
lungs versus volume of air within the lungs over time for each breath.
This plot allows doctors to investigate the relationship between flow
and volume at various points in time during a given breath, which can
indicate whether one has normally functioning lungs, suffers from an
obstructive airway disease (such as asthma), suffers from a restrictive
lung disease (such as certain types of pneumonia), or rather has
another condition altogether. In fact, certain pulmonary ailments are
associated with certain shapes of these loops. Figure~\ref{figdisease}
shows Flow-Volume Loops for healthy patients and for
patients with varying pulmonary ailments [\citet{loop}].

%
%

\citet{cosinor} discuss a frequentist approach for comparing two
functions through the use of a Fourier basis expansion.
Behseta and Kass (\citeyear{barscompare}) propose a Bayesian method for assessing the
equality of two functions using a nonparametric regression method known
as Bayesian adaptive regression splines (BARS). Neither of these
approaches uses the idea of establishing \textit{practical}
equivalence; rather, both papers test strict equality between the
functions of interest, and in fact set strict equality as the null
hypothesis and lack thereof as the alternative. In this paper, we
propose a framework for functional equivalence testing that is
analogous to its univariate counterpart. This involves an extension of
scalar techniques to the functional realm and a modification of said
techniques when a simple extension is not possible. In so doing, the
inferential objective becomes to establish that a \textit{functional}
metric of equivalence lies within a tolerance region with a
prespecified Type I error rate. We then discuss methods for
equivalence testing within the frequentist and Bayesian paradigms, and
illustrate these techniques with data from the method comparison study
between SLP and spirometry. We further introduce a Bayesian model for
heteroscedastic functional data inspired by the work of \citet{bar00}
that separately places priors on the correlation structure and the
underlying variance functions.

\section{A framework for equivalence testing}\label{sec2}

\subsection{Equivalence testing for scalar data}\label{sec21}

In the scalar case, equivalence testing begins by defining a metric
whose value can be used to assess equivalence between the two
populations of interest, say, $\theta$. Common choices include the
difference between group means, $\mu_1 - \mu_2$, and the difference of
logarithms of group means, $\log(\mu_1) - \log(\mu_2)$ (provided one's
data are strictly positive). One then chooses lower and upper
thresholds, $\kappa_l$ and $\kappa_u$, such that we can reject or fail
to reject nonequivalence depending on whether or not $\theta$ falls
between $\kappa_l$ and $\kappa_u$. The null hypothesis is
nonequivalence and the alternative is equivalence:
\begin{eqnarray*}
 && {\mathbf{H}}_{\mathbf{0}}\dvtx  \theta\notin(\kappa_l,
\kappa_u),
\\
&& {\mathbf{H}}_{\mathbf{a}}\dvtx  \theta\in(\kappa_l,
\kappa_u).
\end{eqnarray*}
A common approach for conducting this hypothesis test within the
frequentist paradigm is known as a Two One-Sided
Test (TOST) [\citet{ber96}]. As the name suggests, this is a two step
procedure. In no particular order, one separately tests for the
alternatives that $\theta< \kappa_u$ and $\theta> \kappa_l$ with each
test being conducted with size $\alpha$. If one successfully
rejects for both tests, practical equivalence may then be suggested at
size $\alpha$; otherwise, one fails to suggest practical equivalence.
The lack of compensation in the significance level of the individual
tests (say, to $\alpha/2$) follows immediately from the theory of
Intersection-Union
Tests (or IUTs), which are tests for which the null parameter space can
be described as the union of disjoint sets, and the alternative as the
intersection of the
complements of those sets. One can see that an equivalence test is an
IUT [\citet{ber82}], as its null region is $\Theta_0:= \{(-\infty,\kappa_l] \cup[\kappa_u,\infty)\}$
and its alternative region is $\Theta_a := (\kappa_l, \kappa_u) =
\Theta_0^c$.

The TOST testing procedure can suffer from a lack of power.
\citet{bro95} and \citet{ber96} propose procedures which are
uniformly more powerful for the scalar case, however, these methods are
themselves quite complicated even when dealing with univariate data, to
such an extent that TOST continues to be the method of choice in the
vast majority of applications. We proceed within the TOST framework,
which not only has intuitive appeal but can also be naturally extended
to a test of equivalence for functional data within the frequentist
paradigm.

The most common goal of equivalence testing is to prove equivalence of
means, but this may not be sufficient. \citet{and90} and \citet{liu92}
both suggest that in addition to comparing mean responses, the variance
of the two responses should also be compared, as a device or drug with
smaller variability may be preferred. We will thus include a test for
equivalence of variance in our testing procedure.

\subsection{Equivalence testing for functional data}\label{secequiv}

We now extend the equivalence testing framework to the functional
regime. Let $\theta(\cdot)$ denote a functional measurement of
similarity between the location parameters of two functions. One
potential choice for $\theta(\cdot)$ is the difference between overall
mean functions. $\mu_1(\cdot) - \mu_2(\cdot)$, but the choice of
$\theta(\cdot)$ should depend on the nature of the inference being
conducted. Let $\kappa_l(\cdot)$ and $\kappa_u(\cdot)$ denote lower and
upper equivalence \textit{bands}, which again vary over the same
continuum as do the functional data. These bands are chosen such that
practical equivalence can be suggested or refuted depending on whether
or not $\theta(\cdot)$ falls entirely within $\kappa_l(\cdot)$ and
$\kappa_u(\cdot)$.

For testing the equivalence of variability of the functional data, let
$\lambda(\cdot)$ be a~measurement of similarity between spreads of the
populations. Choices may include
$\frac{\sigma^2_1(\cdot)}{\sigma^2_2(\cdot)}$, the ratio between the
variance functions of the two populations, or $\sigma^2_1(\cdot) -
\sigma^2_2(\cdot)$, the difference between the two variances. We again
establish upper and lower bands, $\zeta_l(\cdot)$ and $\zeta_u(\cdot)$,
within which we can suggest practical equivalence of variance
functions.

The null and alternative hypotheses for the tests of location and
spread can then be stated as follows:
\begin{eqnarray*}
&& {\mathbf{H}}_{\mathbf{0}}^{\bolds{\theta}} \dvtx \exists t\in\mathcal{T} \ni\theta(t)
\notin\bigl(\kappa_l(t), \kappa_u(t)\bigr),
\\
&& {\mathbf{H}}_{\mathbf{a}}^{\bolds{\theta}}\dvtx  \forall t \in\mathcal{T}, \theta(t)
\in\bigl(\kappa_l(t), \kappa_u(t)\bigr),
\\
&& {\mathbf{H}}_{\mathbf{0}}^{\bolds{\lambda}}\dvtx  \exists t\in\mathcal{T} \ni\lambda(t)
\notin\bigl(\zeta_l(t), \zeta_u(t)\bigr),
\\
&& {\mathbf{H}}_{\mathbf{a}}^{\bolds{\lambda}}\dvtx \forall t \in\mathcal{T}, \lambda(t)
\in\bigl(\zeta_l(t), \zeta_u(t)\bigr).
\end{eqnarray*}
Note that the above test, in aggregate, is an IUT; the alternative
space is $\{ \theta(\cdot), \lambda(\cdot)\dvtx \forall t \in\mathcal{T},
\theta(t) \in(\kappa_l(t), \kappa_u(t)) \cap\lambda(t) \in(\zeta
_l(t), \zeta_u(t))\}$. In order to test these hypotheses within the frequentist
paradigm, we propose conducting two TOST procedures, one each for the
location and spread parameters. Since this is an IUT, each of the four total
hypothesis tests can be conducted at significance level $\alpha$ to
arrive at an overall size of $\alpha$. Details of our frequentist
testing procedure
can be found in Section~\ref{secfrequentist}. Sections~\ref{sec8} and~\ref{sec9} also
discuss conducting this test as a Bayesian.

Falling outside of the equivalence region for variability need not be a
condemnation; to the contrary, whichever population has markedly
smaller variability could be favored on those grounds. If one were
comparing a gold standard to a new device and the new device had
markedly lower variation, that would strengthen the case for the
introduction of the new device into the market. Hence, in the case of
method comparison studies, a simple one-sided test of noninferiority
may be sufficient for comparing residual variability.

Note that, in practice, functional data are measured along a finite
grid of values. Thus, the grid must be fine enough such that areas of
potential dissimilarity along the domain are not ignored.

\section{Equivalence testing for volume over time
functions}\label{sec3}
\begin{figure}

\includegraphics{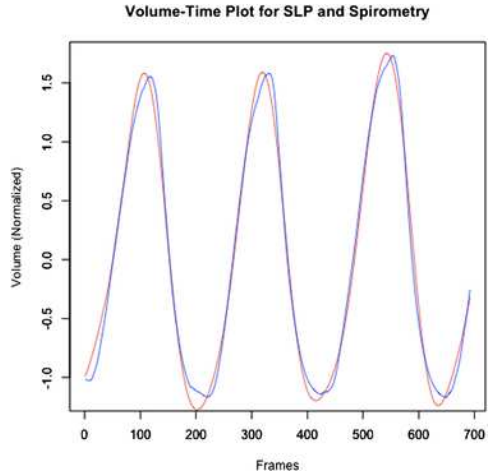}

\caption{Volume over time obtained using SLP and spirometry
for an individual.}\label{figoverlay}
\end{figure}

As was explained in Section~\ref{secintro}, we are interested in
whether or not the Flow-Volume Loops produced by spirometry are
practically equivalent to those produced by SLP in terms of location
and variability. Measurements for volume over time and flow over time
were recorded in 2009 for 16 individuals, with the devices set up such
that each breath was simultaneously recorded by SLP and spirometry.
These data were not the result of a clinical trial and, hence, our use
of the data serves exposition of our methodology rather than an
argument for the equivalence of SLP and spirometry. Our analysis
herein focuses on using the 453 pairs of volume over time curves
measured by both devices on these 16 patients to assess the equivalence
of SLP and spirometry. Figure~\ref{figoverlay} shows the visual
correspondence between these volume over time plots for SLP and
spirometry from an individual.

%
%

The data require preprocessing before our analysis can proceed, as we
must break our recordings into individual breaths that are aligned
between devices and that are comparable in terms of their domains and
scale; see the supplementary materials [\citet{supp}]  for details. This results in 453
pairs of breaths, where each breath is measured at 25 equispaced time
points, time is scaled to the interval $[0,1]$, and time $t$ for SLP
corresponds with time $t$ for spirometry within each pair to the best
of our ability.

\subsection{A model for volume over time functions}\label{secmodel}

We use a functional analysis of variance model with cross-covariance
between pairs of functions for our data. Functional analysis of
variance models are appropriate when one's data are comprised of
functional responses that are believed to differ from one another
solely due to certain categorical variables [\citet{functional}]. Our
model states that we can express the measured volume in the lungs of
person $i$ using both devices (denoting SLP by $1$ and spirometry by
$2$) in the $k$th breath at time $t \in\mathcal{T}$ as follows:
\begin{eqnarray*}
\left[
\begin{array}{c} \mathbf{v}_{i,1,k}(t)
\\
\mathbf{v}_{i,2,k}(t)
\end{array}
\right]
&= &
\left[
\begin{array}{c}
\alpha_{i,1}(t)
\\
\alpha_{i,2}(t)
\end{array}
\right]
+
\left[
\begin{array}{c}
\epsilon_{i,1,k}(t)
\\
\epsilon_{i,2,k}(t)
\end{array}
\right],
\\
\left[
\begin{array}{c}
\alpha_{i,1}(t)
\\
\alpha_{i,2}(t)
\end{array}
\right]
&= &
\left[
\begin{array}{c}
\mu_{1}(t)
\\
\mu_{2}(t)
\end{array}
\right]
+
\left[
\begin{array}{c}
\varepsilon_{i,1}(t)
\\
\varepsilon_{i,2}(t)
\end{array}
\right].
\end{eqnarray*}
In this model $[\mu_1(\cdot), \mu_2(\cdot)]$ represent the overall mean
volume over time trajectory for each device. We model the pairs $\{
[\alpha_{i, 1}(\cdot), \alpha_{i,2}(\cdot)]\}$ as random effects, as
we think of the individuals as draws from a larger population. The
terms $\{[\epsilon_{i,1,k}(\cdot), \epsilon_{i,2,k}(\cdot)]\}$ are the
mean zero error functions for the realized volume over time trajectory
of each pair of devices, assumed to be independent between breaths
while allowing for both strong autocorrelation along the domain of a
given breath and cross-correlation between two breaths in a given pair.
This means that not only is there correlation between the value of the
functions at times $t$ and $t'$ for each breath from a specific device,
but there will also be a correlation between the observation at time
$t$ from SLP and the observation at time $t'$ from the spirometer.
Denote the variance functions of these errors by $[\sigma^2_{\epsilon
,1}(\cdot), \sigma^2_{\epsilon,2}(\cdot)]$. The terms $\{
[\varepsilon
_{i,1}(\cdot), \varepsilon_{i,2}(\cdot)]\}$ are the mean zero error
functions for each patient's pair of random effects, assumed to be
independent between patients while allowing for both strong
autocorrelation along the domain of a given breath and
cross-correlation between random effects in a given pair. Denote the
variance functions of these random effects by $[\sigma^2_{\alpha
,1}(\cdot),\ \sigma^2_{\alpha,2}(\cdot)]$.

\subsection{Defining equivalence bands}\label{secbands}
For\vspace*{1pt} our analysis, we define $\theta(\cdot) \triangleq\mu_1(\cdot) -
\mu_2(\cdot)$, $\lambda(\cdot) \triangleq
\sigma^2_{\epsilon,1}(\cdot)/\sigma^2_{\epsilon,2}(\cdot)$. In
addition, we want to assure ourselves that the variabilities of the
random effect functions are similar between the two populations;
otherwise, there may be evidence of a systematic bias. As such, we
define a third metric of equivalence as $\psi(\cdot) \triangleq
\sigma^2_{\alpha,1}(\cdot)/\sigma^2_{\alpha,2}(\cdot)$. Research is
currently being conducted to ascertain proper values for upper and
lower equivalence bands for our measures of equivalence of location and
spread. These equivalence bands must be established via consultation of
field experts (in our case, with pulmonary specialists). For the
purpose of illustrating the methodology outlined herein, however, we
set reasonable equivalence bands based on the fact that the time points
immediately before, during, and immediately after maximal volume is
attained are critical for diagnostic purposes: $\kappa_l(t) \triangleq
-0.05\cos(2\pi t) - 0.15$; $\kappa_u(t) \triangleq 0.05 \cos(2\pi t) +
0.15$; $\zeta_u(t) \triangleq 0.1\cos(2\pi t) + 1.8$; $\zeta_l(t)
\triangleq 1/(0.1\cos(2\pi t) + 1.8)$.

\begin{figure}

\includegraphics{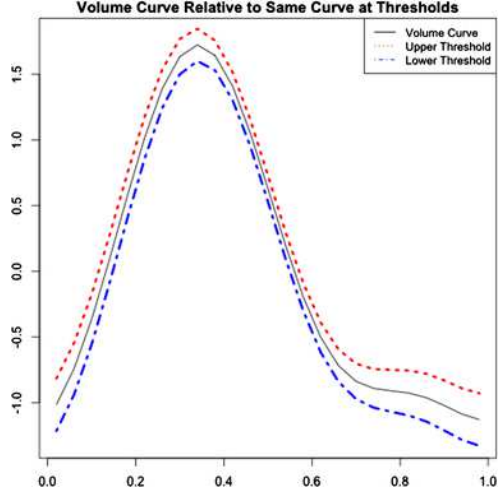}

\caption{A Volume curve and the corresponding curve with $\kappa
_l(\cdot)$ and $\kappa_u(\cdot)$ applied.}
\label{figlocthreshold}
\end{figure}

We use the same sets of equivalence bands for the error variances and
the random effect variances, although in practice these should be
chosen separately. The class of equivalence bands need not be
symmetric, as this assumption may be unrealistic; we have merely done
so for simplicity. Figure~\ref{figlocthreshold} shows the locational
discrepancy between volume curves if the true differences between
devices truly were at the upper and lower thresholds of equivalence we
have specified.

%
%

\section{Frequentist equivalence testing for functional
data}\label{secfrequentist}

We propose using the nonparametric bootstrap [\citet{efr93}] for
assessing equivalence by constructing pointwise confidence intervals
for each metric of equivalence, and then using the duality between
confidence intervals and pointwise hypothesis tests to conduct our
inference. We begin with a testing procedure for i.i.d. data, as we
imagine many situations encountered in practice will be of this form.
We then discuss a procedure for testing within a random effects model.
Allowing for random effects is useful for repeated measures data such
as our pulmonary device data. Through our exposition, we illustrate why
pointwise coverage of our confidence intervals is actually sufficient
for guaranteeing that the resultant inference is of the desired size.

\subsection{IID data, independence between populations}
\label{secindependent}

We use the difference in mean functions, $\theta(\cdot) \triangleq
\mu
_1(\cdot) - \mu_2(\cdot)$ and the
ratio of variance functions $\lambda(\cdot) \triangleq\frac{\sigma
^2_{\epsilon,1}(\cdot)}{\sigma^2_{\epsilon,2}(\cdot)}$,
as metrics for equivalence. Let $y_{1,1}(\cdot),\ldots, y_{1,n_{1}}(\cdot)$
and $y_{2,1}(\cdot),\break  \ldots, y_{2,n_{2}}(\cdot)$ denote
the $n_1$ and $n_2$ observations from groups 1 and 2, respectively,
and let $\bar{y}_1(\cdot) - \bar{y}_2(\cdot)$ denote the sample
mean functions.

We use $\hat{\theta}(\cdot) \triangleq\bar{y}_1(\cdot) -
\bar{y}_2(\cdot)$ and $\hat{\lambda}(\cdot) \triangleq
\frac{s^2_{\epsilon,1}(\cdot)}{s^2_{\epsilon,2}(\cdot)}$ as our test
statistics for the hypothesis test, and use the nonparametric bootstrap
to derive pointwise confidence intervals for the corresponding
parameters. We then use the duality between one-sided confidence
intervals and one-sided tests to reject or fail to reject
nonequivalence.

In each bootstrap simulation, we do the following:
\begin{longlist}[4.]
\item[1.] Sample $n_1$ curves \textit{with replacement} from the curves
in group 1, and sample $n_2$ curves \textit{with replacement}
from the modified curves in group~2.
\item[2.] Compute the pointwise mean curve from these samples and the
pointwise variance curves for each population. Denote these as
$ \{\bar{y}^{*}_i(\cdot) \}$ and
$ \{s^{2*}_i(\cdot) \}$.
\item[3.] Compute $\hat{\theta}^{*}(\cdot) \triangleq
\bar{y}^{*}_1(\cdot) - \bar{y}^{*}_2(\cdot)$ and
$\hat{\lambda}^{*}(\cdot)\triangleq\frac{s^{2*}_1(\cdot
)}{s^{2*}_2(\cdot)}$.
\item[4.] Store this value.
\end{longlist}

Next, we find upper and lower one-sided pointwise $100(1-\alpha)$
confidence intervals. Let $q_p[X(t)]$ denote the $p$-quantile for the
distribution of $X$ evaluated at time $t$. Then, we define our upper
and lower pointwise confidence intervals for $\theta(t)$ using a bias
correcting percentile-based bootstrap as discussed in
\citet{dav97}:
\begin{eqnarray*}
C^u_{1-\alpha}\bigl(\theta(t)\bigr) &=& \bigl[2\hat{\theta}(t) -
q_{\alpha}\bigl[\hat{\theta}^{*}(t)\bigr], \infty\bigr),
\\
C^l_{1-\alpha}\bigl(\theta(t)\bigr) &=& \bigl(-\infty, 2\hat{
\theta}(t) - q_{1-\alpha}\bigl[\hat{\theta}^{*}(t)\bigr]\bigr).
\end{eqnarray*}
At any particular
poin $t$, $C^u_{1-\alpha}(\theta(t))$ and $C^l_{1-\alpha}(\theta(t))$
can be interpreted as the set of all $\theta_0$ such that we fail to
reject the null that $\theta(t) \leq\theta_0$ and $\theta(t) \geq
\theta_0$, respectively. As such, if our lower equivalence band at time
$t$, $\kappa_l(t)$, is outside of $C^u_{1-\alpha}(\theta(t))$, then we
can reject the null that $\theta(t) \leq\kappa_l(t)$ at the point $t$.
Likewise, if $\kappa_u(t)$ is outside of $C^l_{1-\alpha}(\theta(t))$,
then we can reject the null that $\theta(t) \geq\kappa_l(t)$ at the
point $t$.

Our upper and lower pointwise confidence interval for $\lambda(t)$ take
on a different form. This is because
dispersion measures are not typically variance stabilized. In such
cases, conventional bootstrap intervals fail
to attain their advertised coverage probabilities in small samples. We
imagine that most test statistics for
testing equivalence of dispersion will be based on the sample variance.
For many distributions (including the normal),
transforming by the logarithm results in an estimator whose variance is
stabilized. Hence, we instead construct upper and
lower one-sided confidence intervals for the variance stabilized
quantity $\log(\lambda(t))$, and then utilize the monotonicity of the
log transform to result in confidence intervals for~$\lambda(t)$,
\begin{eqnarray*}
C^u_{1-\alpha}\bigl(\lambda(t)\bigr) & =& \bigl[\bigl(\hat{
\lambda}(t)\bigr)^2\times q_{1-\alpha}\bigl[1/\hat{
\lambda}^{*}(t)\bigr], \infty\bigr),
\\
C^l_{1-\alpha}\bigl(\lambda(t)\bigr) &=& \bigl(0, \bigl(\hat{
\lambda}(t)\bigr)^2\times q_{\alpha}\bigl[1/\hat{
\lambda}^{*}(t)\bigr]\bigr].
\end{eqnarray*}
These intervals can
be used to test whether $\lambda(t)$ is below the upper equivalence
band and above the lower equivalence band at any point $t$. If one is
concerned about the log transform providing variance stabilization,
another approach to constructing these confidence intervals would be to
estimate a variance stabilizing transformation within the bootstrap
framework [see \citet{dav97,tib88}].

We now have tests for whether or not we have equivalence of location
and spread at any point $t$. To test for overall equivalence, we
conduct tests at each domain point based on the $100(1-\alpha)$
pointwise interval at all points $t\in\mathcal{T}$ and reject the null
of nonequivalence only if all of the individual tests result in a
rejection. To see why there is no need to correct for simultaneous
comparisons, let $\mathcal{T}_0 \cup\mathcal{T}_a = \mathcal{T}$ be a
partition of the domain where $\mathcal{T}_0$ contains the points for
which the null hypothesis is true and $\mathcal{T}_a$ contains the
points for which the alternative is true for any true metric of
equivalence in the set of nonequivalence. Then, the probability of a
false rejection is bounded as follows:
\begin{eqnarray*}
\mathbb{P}(\mbox{Type I error}) &=& \mathbb{P}(\mbox{falsely reject
all of }\mathcal{T}_0, \mbox{correctly reject all of }\mathcal{T}_a)
\\
&\leq & \mathbb{P}(\mbox{falsely reject all of }\mathcal{T}_0)
\\
&\leq & \mathbb{P}(\mbox{falsely reject a particular } t_0 \in
\mathcal{T}_0)
\\
&= &\alpha.
\end{eqnarray*}
Hence, pointwise $\alpha$ tests of hypothesis
guarantee size of at most $\alpha$. In fact, if one had further
information regarding the correlation between test statistics, these
tests could be done at a size larger than $\alpha$, since our decision
to reject nonequivalence is an intersection of tests. As an example, if
our function were defined on a grid of size $|\mathcal{T}| = 20$, our
test statistics were independent, and we wanted an overall size of
$\alpha= 0.05$, we could then run our tests using $\alpha^* =
\alpha^{1/20} = 0.87$. In the absence of such knowledge, conducting the
pointwise tests at size $\alpha$ is actually a tight upper bound. To
see this, consider an equivalence metric that is in the equivalence
region at all points along the domain except for $t_0$, at which its
value equals that of the equivalence band. If the probabilities of
correct rejection at all points $\mathcal{T}/\{t_0\}$ are sufficiently
close to one, then essentially the type one error rate is the size of
the test at $t_0$, which is $\alpha$. In Section~\ref{secsize}, we
give an example where the overall size approaches the upper bound
$\alpha$.

\subsection{IID matched pairs}
For paired functions (commonly arising in comparison studies where
simultaneous measurements using two devices are possible), slight
alterations are required in the bootstrapping procedure. We again use
the difference in mean functions, $\theta(\cdot) \triangleq
\mu_1(\cdot) - \mu_2(\cdot)$ and the ratio of variance functions
$\lambda(\cdot) \triangleq
\frac{\sigma^2_1(\cdot)}{\sigma^2_2(\cdot)}$, as metrics for
equivalence. Let $\{y_{1,i}(\cdot), y_{2,i}(\cdot)\}$ be the paired
curves, and let $n$ denote the total number of pairs. The bootstrap
procedure is as follows:
\begin{longlist}[4.]
\item[1.] Sample $n$ pairs of curves \textit{with replacement} from the
original sample.
\item[2.] Compute the pointwise mean curve from these samples and the
pointwise variance curves for each population. Denote these as
$[\bar{y}^{*}_1(\cdot), \bar{y}^*_2(\cdot)]$ and
$[s^{2*}_1(\cdot), s^{2*}_2(\cdot)]$.
\item[3.] Compute $\hat{\theta}^{*}(\cdot) \triangleq
\bar{y}^{*}_1(\cdot) - \bar{y}^{*}_2(\cdot)$ and
$\hat{\lambda}^{*}(\cdot)\triangleq\frac{s^{2*}_1(\cdot
)}{s^{2*}_2(\cdot)}$.
\item[4.] Record this value.
\end{longlist}
Now that our bootstrap samples have been acquired, the rest of the
procedure is identical to that explained in Section~\ref{secindependent}.

\subsection{Random effects with matched pairs}
We now describe a nonparametric bootstrap procedure for paired random
effects and paired responses. The procedure for nonmatched data would
replace sampling pairs with sampling individually from two populations
and, hence, we omit its discussion herein. See \citet{cha13} for an
overview of random effect bootstrapping procedures.

Suppose our data consist of $A$ individuals with pairs of random
effects $[\alpha_{i,1}(\cdot), \alpha_{i,2}(\cdot)] \stackrel
{\mathrm{i.i.d.}}{\sim}F$ with mean $[\mu_1(\cdot), \mu_2(\cdot
)]$ and variance
$[\sigma^2_{\alpha,1}(\cdot), \sigma^2_{\alpha,2}(\cdot)]$. For each
individual $i\in[A]$, we observe $n_i$ pairs of curves\vspace*{1pt} with
$[y_{i,1,k}(\cdot),\break  y_{i,2,k}(\cdot)] \stackrel{\mathrm{i.i.d.}}{\sim} G_i$ with
mean $[\alpha_{i, 1}(\cdot), \alpha_{i,2}(\cdot)]$ and variance
$[\sigma^2_{\epsilon,1}(\cdot), \sigma^2_{\epsilon,2}(\cdot)]$.
Let $N
= \sum_{i=1}^A n_i$ denote the total number of curves. Our test for
equivalence will, as before, focus on the location metric
$\theta(\cdot)\triangleq\mu_1(\cdot) - \mu_2(\cdot)$ and metric of
equivalence of error variabilities, $\lambda(\cdot) \triangleq
\frac{\sigma^2_{\epsilon, 1}(\cdot)}{\sigma^2_{\epsilon, 2}(\cdot)}$.
As described in Section~\ref{secbands}, we also include a third
metric, the ratio of random effect variances of the two populations:
$\psi(\cdot) \triangleq
\frac{\sigma^2_{\alpha,1}(\cdot)}{\sigma^2_{\alpha,2}(\cdot)}$.

Let $\bar{y}_{j}(\cdot)\triangleq
\frac{1}{N}\sum_{i=1}^A\sum_{k=1}^{n_{i}} y_{i,j,k}(\cdot)$ be the
overall mean curve for coordinate $j$ and let $\bar{y}_{i,j}
\triangleq\frac{1}{n_i}\sum_{k=1}^{n_{i}} y_{i,j,k}(\cdot)$ be the mean
curve for coordinate $j$ of individual $i$. Now, define $\operatorname{SSE}_j(\cdot)
\triangleq\sum_{i=1}^{A}\sum_{k=1}^{n_{i}}(y_{i,j,k}(\cdot) -
\bar{y}_{j}(\cdot))^2$, and let $\operatorname{SSA}_j(\cdot) \triangleq\sum_{i=1}^{A}
n_i(\bar{y}_{i,j} - \bar{y}_{j})^2$. Our estimators for these metrics
of equivalence will be based on their univariate random effect
counterparts derived via ANOVA. See \citet{sea09} for a description of
methods for univariate random effect analysis. Begin by defining our
estimate of the random effect variance curve by $s^2_{\alpha, j}(\cdot)
= (\operatorname{SSA}_1(\cdot)/(A-1) - \operatorname{SSE}_1(\cdot)/(N-1))/n^*$, $n^* = (N - (\sum
n_i^2)/N)/(A-1)$. Then, we define our test statistics as
$\hat{\lambda}(\cdot) = \frac{\operatorname{SSE}_1(\cdot)}{\operatorname{SSE}_2(\cdot)}$ and
$\hat{\psi}(\cdot) = s^2_{\alpha,1}(\cdot)/s^2_{\alpha,2}(\cdot
)$. Our
estimators for the random effects will be $\hat{\alpha}_{i,j}(\cdot) =
\bar{y}_{i,j}(\cdot)$. Based on these, we estimate our location metric,
$\theta(\cdot)$, by $\hat{\theta}(\cdot) =
\frac{1}{A}\sum_{i=1}^A(\hat{\alpha}_{i,1}(\cdot)-\hat{\alpha
}_{i,2}(\cdot))$.

Denote $r_{i,j,k}(\cdot) = y_{i,j,k}(\cdot) -
\hat{\alpha}_{i,j}(\cdot)$. We then consider these $N$ pairs as a
reservoir from which to draw error functions in the bootstrap
simulation, rather than maintaining a correspondence between random
effects and residuals from that random effect's group. This ignores the
sample covariance between residuals from the same group and slight
heteroscedasticity if the design is unbalanced. We doubt that this
would have a substantial impact on the inference being performed (which
the simulation studies of Section~\ref{seccompare} seem to suggest),
but leave a proper investigation for future work.

Before beginning the bootstrap, we adjust our estimates of the random
effects such that the ratio of the variances of the pool of random
effects used in the bootstrap matches up with our estimate of the
random effect variance. We define the following adjusted random
effects:
\[
\hat{a}_{i,j}(\cdot) = \bar{y}_{j}(\cdot) - \bigl(\hat{
\alpha}_{i,j}(\cdot) - \bar{y}_{j}(\cdot)\bigr)
\frac{s_{\alpha,j}(\cdot)}{\operatorname{SD}(\hat
{\alpha}_{i,j}(\cdot))}.
\]
Here, $\operatorname{SD}(\hat{\alpha}_{i,j}(\cdot))$ is the standard deviation of our
estimated group means evaluated pointwise. This transformation
guarantees that the variances of the random effects used in the
bootstrap are the same as our estimate of that variance. As noted in
\citet{sha95} and \citet{cha13}, this step is required to assure that
the confidence intervals produced by the bootstrap procedure are
consistent. We then proceed as follows:
\begin{longlist}[5.]
\item[1.] Sample $A$ pairs of random effects from
$\{[\hat{a}_{i,1}(\cdot), \hat{a}_{i,2}(\cdot)]\}$ with
replacement. Call them $\{[\hat{a}_{i,1}(\cdot),
\hat{a}_{i,2}(\cdot)]^{*}\}$. The first pair drawn gets
assigned $n_1$ as the number of pairs of curves to be drawn
within that group, the second gets assigned $n_2$, etc.
\item[2.] For each $i$, draw $n_i$ pairs of residuals with replacement
from $\{[r_{i,1,k}(\cdot),\break  r_{i,2,k}(\cdot)]\}$. Call these
$\{[r_{i,1,k}(\cdot), r_{i,2,k}(\cdot)]^*\}$.
\item[3.] Define $[y_{i,1,k}(\cdot), y_{i,2,k}(\cdot)]^* =
[\hat{a}_{i,1}(\cdot), \hat{a}_{i,2}(\cdot)]^{*}+
[r_{i,1,k}(\cdot), r_{i,2,k}(\cdot)]^*$.
\item[4.] Estimate\vspace*{1pt} $\bar{y}_j^{*}(\cdot), \bar{y}_{i,j}^{*}(\cdot),
\operatorname{SSE}_j^{*}(\cdot), \operatorname{SSA}_j^{*}(\cdot)$ based on the bootstrap
sample $\{[y_{i,1,k}(\cdot), y_{i,2,k}(\cdot)]^*\}$.
\item[5.] Estimate $\hat{\theta}^*(\cdot), \hat{\lambda}^*(\cdot),
\hat{\psi}^*(\cdot)$ based on these quantities.
\end{longlist}

We can create pointwise $100(1-\alpha)$ confidence intervals for
$\theta
(\cdot)$ and $\lambda(\cdot)$ just as we did in Section~\ref{secindependent}. For $\psi(\cdot)$, we define our confidence
intervals in the same manner as we did with $\lambda(\cdot)$,
\begin{eqnarray*}
C^u_{1-\alpha}\bigl(\psi(t)\bigr) &= & \bigl[\bigl(\hat{\psi}(t)
\bigr)^2\times q_{1-\alpha}\bigl[1/\hat{\psi}^{*}(t)
\bigr], \infty\bigl),
\\
C^l_{1-\alpha}\bigl(\psi(t)\bigr) &=&  \bigl(0, \bigl(\hat{\psi}(t)
\bigr)^2\times q_{\alpha}\bigl[1/\hat{\psi}^{*}(t)
\bigr]\bigr].
\end{eqnarray*}
As before, these confidence intervals can be used to test whether $\psi
(t)$ is below the upper equivalence band and above the lower
equivalence band at any point $t$.

%
%
\begin{figure}

\includegraphics{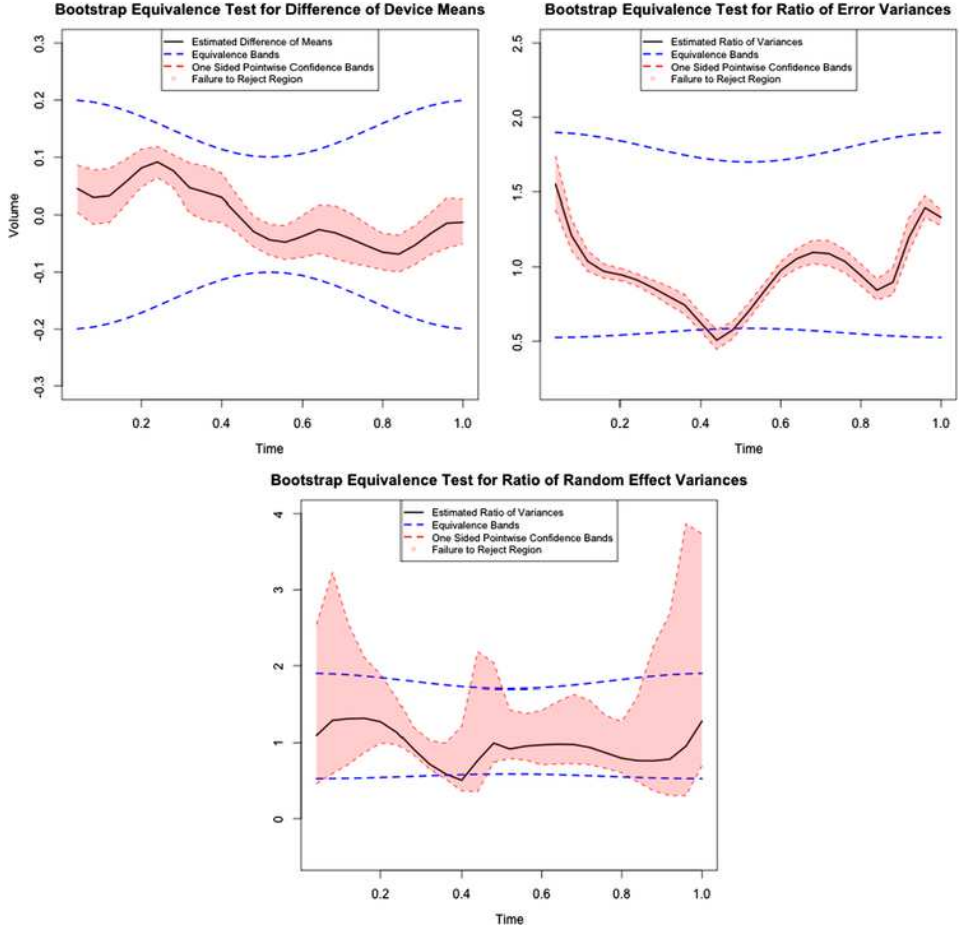}

\caption{Equivalence Test for the difference of means (top left),
ratio of error variances (top right), and the ratio of random effect
variance curves (bottom).}\label{figbootlocation}
\end{figure}

\section{A frequentist test of equivalence for lung volume functional data}
We now conduct our equivalence test using the methods described in
Section~\ref{secfrequentist} for paired random effects. We drew 10,000
bootstrap samples and used $\alpha= 0.05$ to carry out these tests. We
find that Figure~\ref{figbootlocation} is a powerful visual display of
the results of this TOST procedure. In each plot, we display the upper\vspace*{1pt}
and lower equivalence bands. We also display the upper band of the
region $C^l_{0.95}(\cdot)$ and the lower band of the region
$C^u_{0.95}(\cdot)$. Recall that we can reject the null if the upper
equivalence band lies entirely outside the region $C^l_{0.95}(\cdot)$
and if the lower equivalence band lies entirely outside the region
$C^u_{0.95}(\cdot)$. Hence, it is sufficient to check whether or not
either the upper or lower equivalence band at any point intersect the
region defined by the overlap of the two one-sided confidence regions,
which is shaded in the plots. Intersection implies failure to reject,
and lack thereof implies rejection of nonequivalence in favor of
equivalence.

Based on Figure~\ref{figbootlocation}, we conclude that we can suggest
equivalence for our locational metric, but fail to reject the null of
nonequivalence for variability of both errors and random effects. We
believe it will always be the case that a two-sided test for the
variability of random effects is appropriate, as deviations in either
direction indicate substantial differences in the distribution of the
individual level mean curves; however, for certain applications (ours
included), lower error variance will be strictly preferred. If we thus
restrict ourselves to only having the ratio of error variances below
the upper equivalence threshold, then we would also reject the null of
noninferiority of error variability. Note that there does appear to be
an inflation of error variance by a factor of 1.5 at the beginning of
each breath for SLP relative to spirometry. Though the ratio between
the two variances is high at this point, the actual magnitude of the
variances at the beginning of these curves is extremely small for both
devices, which results in the high value for the ratio of variances.

\section{A Bayesian paradigm for equivalence testing}\label{secbayesparadigm}

As in the frequentist case, we suggest using functional measures of
location and spread to assess practical equivalence, however, carrying
out a TOST hypothesis test is not required within the Bayesian
paradigm. Rather than conducting a stochastic proof by contradiction,
the Bayesian paradigm allows us to directly compute posterior
probabilities of our functional metrics of equivalence falling entirely
within specified equivalence ranges. That is, the Bayesian paradigm
allows for direct computation of
$\mathbb{P}\{H_a| \mathrm{Data}\}$ for each of the equivalence hypotheses. In
light of
this, we propose that the researcher conduct the following three steps
when using the Bayesian framework for equivalence testing:
\begin{longlist}[3.]
\item[1.] Define an equivalence region through expert consultation.
\item[2.] Define a probability value, call it $\gamma$, such that if
$\mathbb{P}\{H_a| \mathrm{Data}\} \geq\gamma$, equivalence may be suggested. Using the
suggestions of \citet{jef61} and \citet{kas95}, a value of $\gamma=
0.75$ or $\gamma= 0.95$ may be appropriate.
\item[3.] Specify prior distributions for the metrics of equivalence that
are commensurate with the researcher's prior belief of the alternative
being true relative to the null.
\end{longlist}

The specifics of this implementation depend on the types of prior
distributions used to model the parameters and data. In Section~\ref{secbayesmethod} we discuss the use of Gaussian Processes in modeling
both our data and parameters and describe a model that allows for
specification of priors and posterior inference for our metrics of
equivalence. Though Gaussian Processes are a rich and flexible class of
distributions for functional data, a valuable extension of our work
would be conducting Bayesian equivalence testing for functional data
using nonparametric models.

\section{Bayesian functional equivalence testing for lung volume data}
\label{secbayesmethod}
Kaufman and Sain (\citeyear{kau10}) discuss using functional ANOVA modeling within the
Bayesian paradigm. They begin by assuming that the functional data are
realizations of an underlying Gaussian process with a mean function
depending on the factor levels and a covariance function that describes
the dependence between points along the function's domain. They further
assume that the covariance between errors can be aptly specified as a
member of the class of Mat\'ern covariance functions [\citet{matern}].
The specification of a correlation function works to impose smoothness
between estimated function values and to allow for interpolation at
unobserved domain values. Gaussian process priors with Mat\'ern
covariance functions are used for the mean functions themselves, which
allows for the incorporation of a priori beliefs about both smoothness
and location.

The assumption of homoscedastic variances along the function's domain
is problematic for us, as allowing the error and random effect
variances to change with time is vital to our investigation of
equivalence. We consider a more flexible class of covariance and
cross-covariance functions: $\mathbf{V}_{i,j}(t,t') = \sigma_{i}(t)
\sigma_{j}(t') \mathbf{R}_{i,j}(t,t')$. Here, $\sigma_{\epsilon,i}(t)$
is the error standard deviation function for device $j$ evaluated at
time $t$, and $\mathbf{R}_{\epsilon,i,j}(t,t')$ is either the
correlation function for device $j$ for observations at times $t$ and
$t'$ if $i=j$ or the cross-correlation function between the error at
time $t$ for device $i$ and the error at time $t'$ for device $j$ if
$i\neq j$.

To simplify notation, let $\Xi$ denote the set containing all of our
parameters. Then, we can write our Multivariate Gaussian Process model
for our responses:
\[
\left.\left[
\begin{array}{c} \mathbf{v}_{i,1,k}(\cdot)
\\
\mathbf{v}_{i,2,k}(\cdot)
\end{array}
\right]
\right| \Xi \stackrel{\mathrm{indep}}{\sim}
\operatorname{MVGP} \left(
\left[
\begin{array}{c} \alpha_{i,1} (\cdot)
\\
\alpha_{i,2} (\cdot)
\end{array}
\right],
\left[
\begin{array}{c@{\quad}c}
\mathbf{V}_{\epsilon,1, 1}(\cdot,\cdot)&
\mathbf{V}_{\epsilon,1,2}(\cdot,\cdot)
\\
\mathbf{V}_{\epsilon, 1, 2}(\cdot,\cdot) &
\mathbf{V}_{\epsilon,2, 2}(\cdot,\cdot)
\end{array}
\right]
\right).
\]
Note that, in practice, our response functions are measured only at a
predetermined set of grid points, $\mathbf{t}= \{t_1,\ldots,t_T\}\subset
\mathcal
{T}$. To distinguish this, let the notation $[\mathbf
{v}_{i,1,k}(\mathbf{t}), \mathbf{v}
_{i,2,k}(\mathbf{t})]$ represent the vector whose coordinates are the response
as measured at each of the $T$ grid points, and let the analogous
notation hold for the functional parameters of our models. Hence,
$[\mathbf{v}
_{i,1,k}(\mathbf{t}), \mathbf{v}_{i,2,k}(\mathbf{t})]'$ represents a
$2T\times1$
vector. Using the decomposition proposed in \citet{bar00}, our
covariance functions evaluated at $\mathbf{t}$ can be described in matrix
notation as $\mathbf{V}_{\epsilon,i,j}(\mathbf{t},\mathbf{t})
\triangleq \operatorname{Diag}
(\sigma_{\epsilon,i}(\mathbf{t}) ) \mathbf{R}_{\epsilon
,i,j}(\mathbf{t},\mathbf{t})
\operatorname{Diag} (\sigma_{\epsilon,j}(\mathbf{t}) )$, where $\operatorname{Diag} (\sigma
_{\epsilon,j}(\mathbf{t}) )$ denotes a $T\times T$ matrix whose diagonal
elements are $\sigma_{\epsilon,j}(\mathbf{t})$.

Our assumption of a Multivariate Gaussian Process results in $[\mathbf{v}
_{i,1,k}(\mathbf{t}),\break  \mathbf{v}_{i,2,k}(\mathbf{t})]$ following a
Multivariate Normal
distribution when we consider observations at the set of gridpoints
$\mathbf{t}$ with the fixed grid analogues for the mean and covariance structure.

\section{Bayesian methodology}\label{sec8}
\subsection{Correlation structure}\label{seccorrelation}
Our data set consists of a total of 453 breaths collected from 16
individuals, where each breath was measured at 25 equispaced time
points using both SLP and spirometry. Our desire to model
cross-covariances between devices results in our matrices of
observations being 50 dimensional. For modeling the error correlation,
this is not an issue, as we have 453 observations, however, as we only
have 16 individuals, a simplifying assumption must be made to proceed.
In many functional data settings, the goal of the data analysis is mean
function estimation and prediction at new locations (kriging). To
facilitate this, modelers typically restrict themselves to a particular
class of correlation functions. Unfortunately, the distribution of
posterior variance functions is highly dependent on the correlation
structure. Hence, misspecification of the correlation model can result
in estimates for variance parameters that are biased and wildly
misleading. As we would like to conduct inference for the ratio of
variance functions of both errors and random effects, we are left
searching for an alternative. More advanced methods that make no
assumptions on the correlation function class have been suggested in
the geostatistics literature [see \citet{nyc02,pac06,fue01,fue02}]
and elsewhere [see \citet{mor06,che12}], but none of these works have
directly focused on the accuracy of the resultant variance estimates.
Estimation of correlation functions for repeatedly observed functional
data remains an active area of research, particularly in the regime
where the number of functional observations is small relative to the
grid size.

Our recommendation is that if the researcher has sufficient data to
flexibly model the correlation structure of both the random effects and
the errors, then this should be the course pursued. As we do not, we
instead make a modeling decision that will facilitate valid inference
for our variance functions. We assume the following structure for the
correlation of our errors and random effects:
\[
\mathbf{R}_{i,j} \bigl(t,t' \bigr) =
\cases{ 1, &\quad $i = j, t = t'$,
\cr
\rho(t), &\quad $i\neq j, t = t'$,
\cr
0, & \quad\mbox{otherwise}.}
\]
We thus primarily focus on the marginal distributions for estimation of
our mean functions and variances. This has the obvious drawback of not
fully exploiting the functional nature of our data, but allows for
estimation of marginal variances without the risk of biases due to
misspecification of the correlation structure. This is an interesting
instance where the simplifying assumptions made to facilitate inference
would not necessarily align with ones made if the goal was estimation
of mean functions or prediction of values at unmeasured locations. In
the latter case, one would likely enforce a restriction to a specific
class of correlation functions which would result in both smooth curve
estimates and a principled manner by which interpolation and prediction
could be performed; however, this would result in misleading estimates
for the variance components of the model, which is unacceptable for
testing equivalence of variance functions. In Section~\ref{seccompare}
we investigate the ramifications of this modeling decision on the
resultant inference.

\subsection{Prior distributions} \label{secprior}
Specification of priors for $\sigma_{\epsilon,1}(\cdot)$ and $\sigma
_{\epsilon,2}(\cdot)$ must be done carefully, as practical
equivalence of error variability is tested using a function of these
parameters. We model these functions as themselves being realizations
of independent stochastic processes. Specifically, we extend the work
of \citet{bar00} to the functional regime by modeling the standard
deviation curves as emanating from Log-Gaussian Processes:
\begin{eqnarray*}
\log\bigl(\sigma^2_{\epsilon, 1}(\cdot) \bigr) &\sim &  \operatorname{GP} \bigl(
\tau_\epsilon(\cdot), s^2_\epsilon\Gamma_{\epsilon}(
\cdot,\cdot) \bigr),
\\
\log\bigl(\sigma^2_{\epsilon, 2}(\cdot) \bigr) &\sim  & \operatorname{GP} \bigl
(\tau
^2_\epsilon(\cdot) - \delta_\epsilon(\cdot),
s^2_\epsilon\Gamma_{\epsilon}(\cdot,\cdot) \bigr),
\\
\log\bigl(\sigma_{\epsilon, 1}(\cdot) \bigr) &\indep & \log\bigl
(\sigma
_{\epsilon, 2}(\cdot) \bigr),
\\
p\bigl(\tau_\epsilon(\cdot)\bigr) &\propto & 1,
\\
\delta_\epsilon(\cdot) &\sim & \tfrac{1}{2}\mathbh{1}\bigl\{
\delta_\epsilon(\cdot) = \log\bigl(\zeta_l(\cdot)\bigr)\bigr\}
+ \tfrac{1}{2}\mathbh{1}\bigl\{\delta_\epsilon(\cdot) =\log\bigl(
\zeta_u(\cdot)\bigr)\bigr\},
\end{eqnarray*}
where ${\Gamma}_{\epsilon}(\cdot,\cdot) = \frac
{1}{2}(|t-t'|/a_\epsilon)^2\mathcal{K}_2(d(t,t')/a_\epsilon)$ is a
standard Mat\'ern correlation function [\citet{matern}] with smoothness
parameter $\nu= 2$.

We use the ratio $\frac{\sigma^2_{\epsilon,1}(\cdot)}{\sigma
^2_{\epsilon,2}(\cdot)}$ as our comparative measure for the error
variability of the two devices. Our prior on the standard deviations
yields the following prior for this ratio:
\begin{eqnarray*}
\frac{\sigma^2_{\epsilon,1}(\cdot)}{\sigma^2_{\epsilon
,2}(\cdot)}  &\sim & \frac{1}{2} \bigl(\mbox{Log-GP} \bigl(\log\bigl(
\zeta_l(\cdot)\bigr), 2s^2_{\epsilon}
R_{\sigma}(\cdot,\cdot) \bigr) \bigr) \\
&&{}+ \frac
{1}{2} \bigl( \mbox{Log-GP}
\bigl(\log\bigl(\zeta_u(\cdot)\bigr), 2s^2_{\epsilon}
R_{\sigma
}(\cdot,\cdot) \bigr) \bigr).
\end{eqnarray*}
This is a $50/50$ mixture of two Log-Gaussian Processes with medians at
the upper and lower equivalence thresholds respectively. Hence, we can
place prior probabilities on falling within the equivalence region by
careful choices of $s^2_\epsilon\Gamma_{\epsilon}(\cdot,\cdot)$.
Borrowing from the frequentist paradigm in which it is incumbent upon
the researcher to prove his or her hypothesis beyond a reasonable
doubt, we set the values of these hyperparameters such that the a
priori probability of equivalence is quite small. We set $s_\epsilon^2
=5$ and $a_\epsilon= 0.1$, which results in a prior probability of
falling entirely within the equivalence region of $\mathbb{P}\{\sigma
^2_{\epsilon,1}(\mathbf{t})/\sigma^2_{\epsilon,2}(\mathbf{t}) \in
(\zeta_l(\mathbf{t}),
\zeta_u(\mathbf{t}))\} \approx5\times10^{-8}$.

For the correlations resulting from the paired nature of our data, we
set $\rho_\epsilon(t) \sim\mathcal{U}[-1,1]$ for all $t$.

For our random effects, $\{[\alpha_{i,1}(\cdot), \alpha_{i,2}(\cdot)]\}$, we
use a Hierarchical Gaussian Process prior:
\[
\left[
\begin{array}{c}\alpha_{i,1}(\cdot)
\\
\alpha_{i,2}(\cdot)
\end{array}
\right]
\stackrel{\mathrm{i.i.d.}}{\sim} \operatorname{GP}
\left(
\left[
\begin{array}{c}\mu_1(\cdot)
\\
\mu_2(\cdot)
\end{array}
\right],
\left[
\begin{array}{c@{\quad}c}
\mathbf{V}_{\alpha,1, 1}(\cdot,\cdot)& \mathbf{V}_{\alpha
,1,2}(\cdot,\cdot)
\\
\mathbf{V}_{\alpha,1,2
}(\cdot,\cdot) &\mathbf{V}_{\alpha,2, 2}(\cdot,\cdot)
\end{array}
\right]
\right).
\]

The priors on the variance functions of our random effects, $[\sigma
^2_{\alpha,1}(\cdot), \sigma^2_{\alpha,2}(\cdot)]$, and the correlation
structure are identical to the one used for the error variances.

The posterior distribution for the difference between the device
specific curves, $\mu_{1}(\cdot) - \mu_{2}(\cdot)$, is of
interest for assessing locational equivalence. Thus, proper attention
must be paid to the prior placed on $ \{\mu_1(\cdot), \mu_2(\cdot)\}$
such that the prior does not unduly force the posterior distribution
toward the prespecified equivalence region. Our priors for $\mu
_1(\cdot
)$ and $\mu_2(\cdot)$ are as follows:
\begin{eqnarray*}
\mu_1(\cdot) & \sim& \operatorname{GP}\bigl(\mu_0(\cdot),
s^2_\mu\Gamma_\mu(\cdot, \cdot)\bigr),
\\
\mu_2(\cdot) &\sim& \operatorname{GP}\bigl(\mu_0(\cdot) -
\delta_\mu(\cdot), s^2_\mu\Gamma
_\mu(\cdot, \cdot)\bigr),
\\
\mu_1(\cdot) & \indep& \mu_2(\cdot),
\\
p\bigl(\mu_0(\cdot)\bigr) &\propto& 1,
\\
\delta_\mu(\cdot) &\sim& \tfrac{1}{2}\mathbh{1}\bigl\{
\delta_\mu(\cdot) = \kappa_l(\cdot)\bigr\} +
\tfrac{1}{2}\mathbh{1}\bigl\{\delta_\mu(\cdot) =\kappa
_u(\cdot)\bigr\},
\end{eqnarray*}
where $\Gamma_\mu(t, t')$ is a Mat\'ern correlation function with
smoothness parameter $\nu= 2$. This then implies that our difference
of means has the following prior:
\[
\mu_1(\cdot) - \mu_2(\cdot) \sim\tfrac{1}{2}
\bigl(\operatorname{GP}\bigl(\kappa_l(\cdot), 2s^2_\mu
\Gamma_\mu(\cdot, \cdot)\bigr) \bigr) + \tfrac
{1}{2} \bigl(\operatorname{GP}
\bigl(\kappa_u(\cdot), 2s^2_\mu
\Gamma_\mu(\cdot, \cdot)\bigr) \bigr).
\]
In other words, our prior on the difference in device means is a
$50/50$
mixture of two Gaussian Processes, with means at the upper and lower
equivalence thresholds respectively. We choose a prior that places 1\%
likelihood in the equivalence region and the remaining 99\% outside of
it. To achieve this, we fixed a value of $a_\mu= 0.3$, and then used
the \texttt{uniroot()} and \texttt{pmvnorm()} functions in \texttt{R}
[\citet{R11}] to solve for the value of $s^2_\mu$ such that $\mathbb
{P}\{\mu
_{1}(\mathbf{t}) - \mu_{2}(\mathbf{t}) \in(\kappa_l(\mathbf{t}),
\kappa_u(\mathbf{t}))\} =
0.01$. This value was found to be 0.1. Note that if one has a sense of
an appropriate basis for the mean functions, one could place a prior
$\mu_0(\cdot)\sim\mathcal{N}(\sum a_k\phi_k(\cdot), \sigma^2_\mu
)$ instead of
$p(\mu_0(\cdot))\propto1$. This could allow for regularization of the
functional fits based on this basis while not restricting them to
entirely follow said basis, and would still facilitate our strategy of
putting priors on equivalence commensurate with prior knowledge.

\subsection{Posterior sampling}
\label{secgibbs}

Before conducting inference based on our model specification, we must
devise a sampling schema for the posterior distribution of our
parameters. We use a Metropolis-within-Gibbs sampling algorithm; see
the supplementary materials [\citet{supp}] for details.

%
%
\begin{figure}

\includegraphics{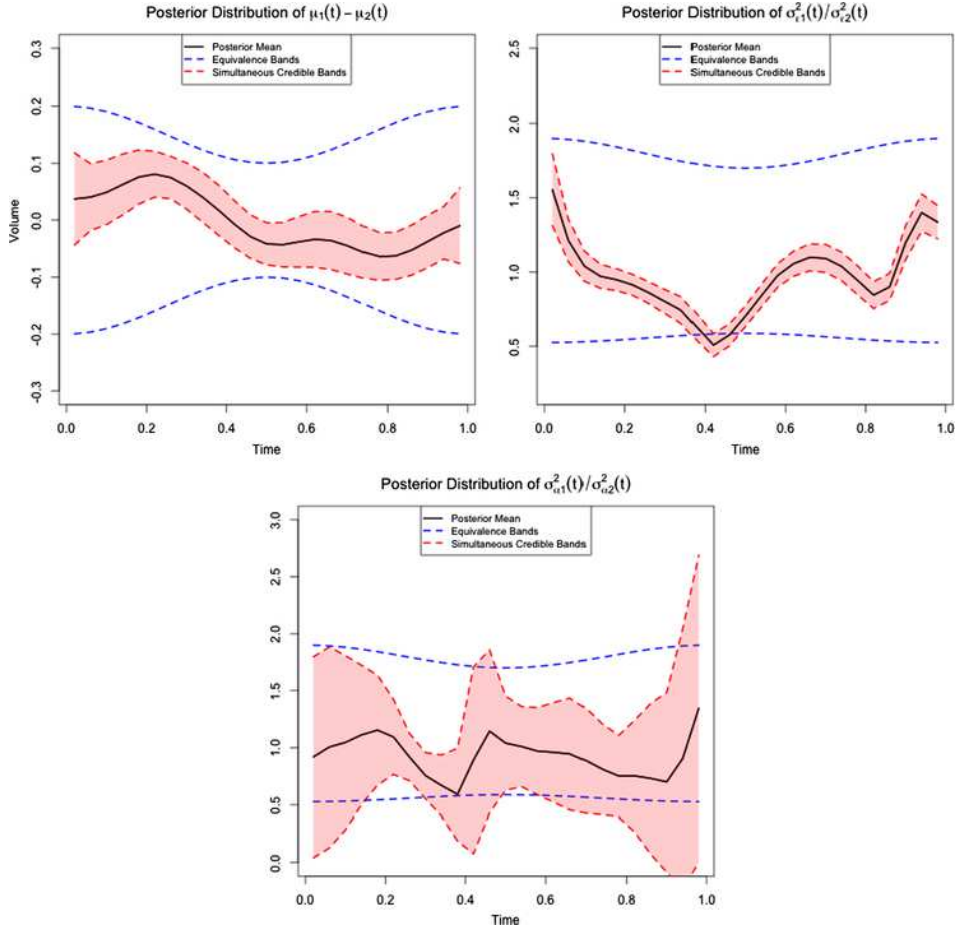}

\caption{95\% simultaneous credible bands for $\mu_1(\cdot) - \mu
_2(\cdot)$ (top left),
$\sigma^2_{\epsilon,1}(\cdot)/\sigma^2_{\epsilon,2}(\cdot)$ (top right),
and $\sigma^2_{\alpha,1}(\cdot)/\sigma^2_{\alpha,2}(\cdot)$ (bottom),
along with upper and lower equivalence bands.} \label{figdeltapost}
\end{figure}

\section{Posterior analysis}\label{sec9}

To conduct our posterior analysis, we ran our Gibbs sampler from three
distinct starting values for 10{,}500 iterations per starting value (for
a total of 31{,}500 iterations). We discarded the first 500 iterations as
burn-in for each chain and took every 10 samples thenceforth for a
total of 1000 samples per starting value, which were then chained
together, resulting in 3000 roughly independent samples. See the
supplementary material [\citet{supp}] for convergence diagnostics.

Figure~\ref{figdeltapost} shows the posterior distribution for the
three metrics of interest. We summarize the posterior distributions of
our metrics of equivalence by the posterior mean curve and 95\%
simultaneous posterior bands. These bands are computed using the
multiplier based method of \citet{buj03}. The posterior bands are
unnecessary for inference, as the computation of $\mathbb{P}\{
H_a|\mathrm{Data}\}$
depends solely on how many posterior curves fall within the equivalence
region, but nonetheless provide a useful graphical aid. For our
locational metric, $\mu_1(\cdot) - \mu_2(\cdot)$, we found that all
3000 of our samples from the posterior distribution fell within the
prespecified equivalence range, suggesting overwhelming evidence in
favor of the hypothesis that these two curves, in terms of location,
can be considered practically equivalent. For the ratio of error
variances, $\sigma^2_{\epsilon,1}(\cdot)/\sigma^2_{\epsilon
,2}(\cdot)$,
we note that if it is the case that lower variability is strictly more
desirable, then 2998 out of 3000 samples fall strictly below the upper
equivalence band; however, if one desires adherence to the lower
equivalence band as well, then our posterior probability of equivalence
is $0.0007$, since our posterior bands regularly violate the lower
tolerance threshold toward the middle of the breaths (around $t=0.5$).
For the ratio of random effect variances, $\sigma^2_{\alpha,1}(\cdot
)/\sigma^2_{\alpha,2}(\cdot)$, we note that although the posterior
median falls well within the equivalence range, only 18.2\% of the
posterior samples fell entirely within the equivalence region. Hence,
although we can suggest equivalence of both means and error variances,
we lack sufficient power to suggest equivalence of random effect variances.

\section{Comparing the frequentist and Bayesian methods}\label{seccompare}
We have presented methods for equivalence testing within the
frequentist and Bayesian paradigms. From a pragmatic perspective, the
relative computational intensity of both methods is of interest to
practitioners. In this respect, our frequentist method is dominant, as
within each bootstrap iteration, only simple vector operations are
required. The Bayesian approach requires sampling from multivariate
distributions, matrix multiplication, matrix inversion, and determinant
calculation within each step. Furthermore, thinning of one out of every
10 iterations was required. Hence, to get the same effective sample
size, we needed to do 10 times as many iterations for the frequentist
procedure as we did for the Bayesian one. To attain 1000 independent
samples via the Bayesian methodology, we needed to run 10{,}500 iterations
of our sampling algorithm, which took 22.6 minutes on a personal laptop
with 4~GB RAM and a 2.7 GHz processor. The bootstrap procedure took
16.1 seconds to run 1000 iterations on the same laptop. This
discrepancy will only increase as the granularity of the grid the user
implements increases, as both determinant and inverse calculation are
$O(p^3)$ in their simplest implementation.

Frequentist and Bayesian inference are not coherent with one another,
in that frequentist inference has a built in preference for the null
hypothesis. For the frequentist, the null is the status quo, and the
goal of the inference is to refute it via a ``proof by contradiction.''
The Bayesian framework, on the other hand, allows the user to put
varying degrees of a priori preference on one hypothesis versus the
other. In our Bayesian analysis we have placed heavy preference on the
null and thus require very strong evidence from the data to put the
posterior probability in the proper region, but this may not always be
appropriate. The Bayesian paradigm allows for a principled manner for
incorporating the results of past studies in the form of the priors
placed on equivalence vs nonequivalence, a feature not offered by the
frequentist framework.

With these caveats in mind, we investigate the size and power of our
methodologies, using the threshold of $\alpha= 0.05$ in the
frequentist procedure. For our Bayesian procedure, we use $\gamma=
0.95$ as our threshold for the posterior probability of equivalence. In
our investigation, we continue to place heavy a priori preference on
nonequivalence for our Bayesian methodology.

\subsection{Type I error}\label{secsize}
\begin{figure}

\includegraphics{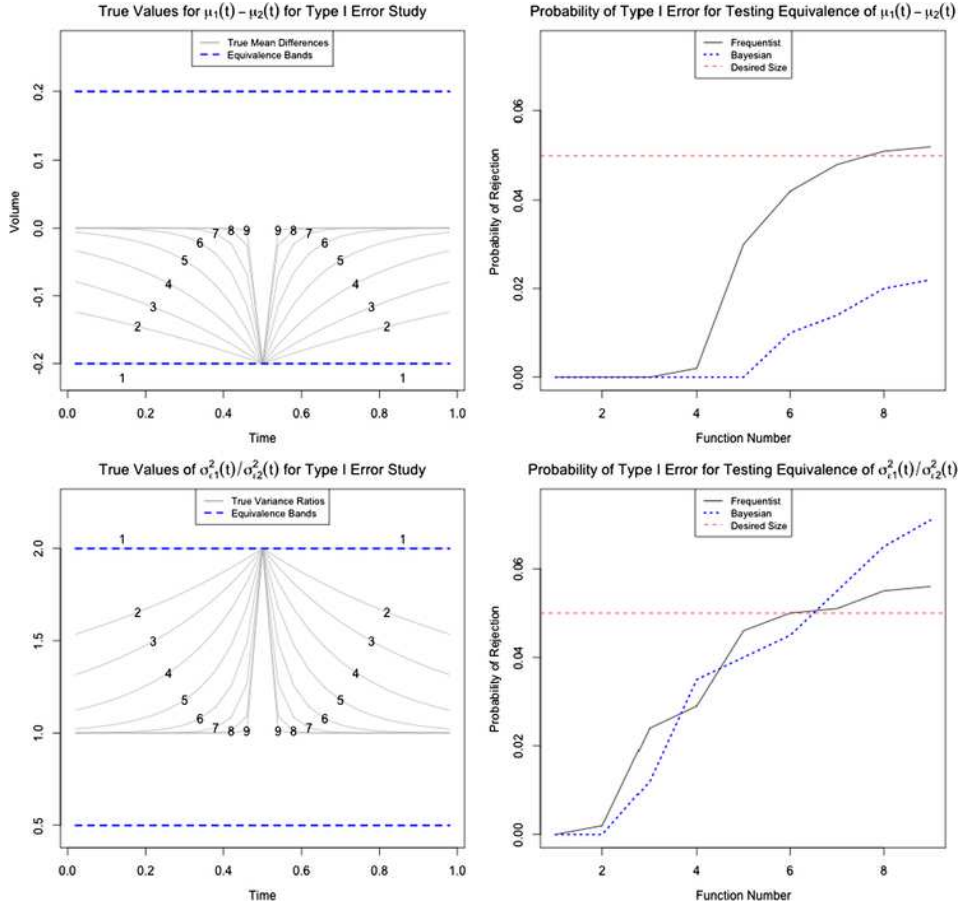}

\caption{Sequence of true values and corresponding Type I error rates
for $\mu_1(\cdot) - \mu_2(\cdot)$
(top) and $\sigma^2_{\epsilon,1}(\cdot)/\sigma^2_{\epsilon
,2}(\cdot)$
(bottom) along with upper and lower equivalence bands used for
Type I error study.} \label{figsizesequence}
\end{figure}

We restrict our investigation to the Type I error rates of our tests
for location and error variances. We simulate 20 matched pair random
effects, and then simulate 20 matched functional responses for each
subpopulation. This results in 400 breaths total. To investigate the
true size of our methods, we define a sequence of true values for our
metrics of equivalence where equivalence is violated at one point along
the domain, and the other points move farther and farther into the
equivalence region. These sequences and numerical labels are shown in
Figure~\ref{figsizesequence}. The remaining values of parameters
needed for simulation are based on the posterior means from our data
set. Additionally, we used an estimate of the correlation structure of
our error functions as the true correlation for simulating both error
functions and random effect functions. This allows us to assess the
robustness of our Bayesian procedure to the assumption of Section~\ref{seccorrelation}

For each of the nine function values in the sequence, we simulated 500
data sets and ran both the frequentist and Bayesian methdologies on
them. Figure~\ref{figsizesequence} shows the result of this study. We
see that for testing the equivalence of mean functions, the Bayesian
procedure is far more conservative than our frequentist procedure,
which appears to be due to the assumption on the correlation structure
made in our Bayesian procedure. As expected, the frequentist procedure
is initially conservative, but has size that approaches 0.05 as the
test becomes increasingly reliant on our data's behavior at one domain
point (the one at which equivalence is violated). Figure~\ref{figsizesequence} also demonstrates that the test is roughly unbiased
in terms of purported size. For testing the equivalence of variances,
the Bayesian and frequentist procedures initially exhibit similar Type
I error rates, and also both appear to be slighty anti-conservative;
however, the Bayesian procedure is anti-conservative to a far more
egregious degree by the end of the sequence of functions, having an
estimated size of 0.072 for the 9th function in the sequence
versus an estimated size of 0.056 for the frequentist procedure at this
value for the true ratio of error variances.
%
%

\subsection{Power}
\begin{figure}

\includegraphics{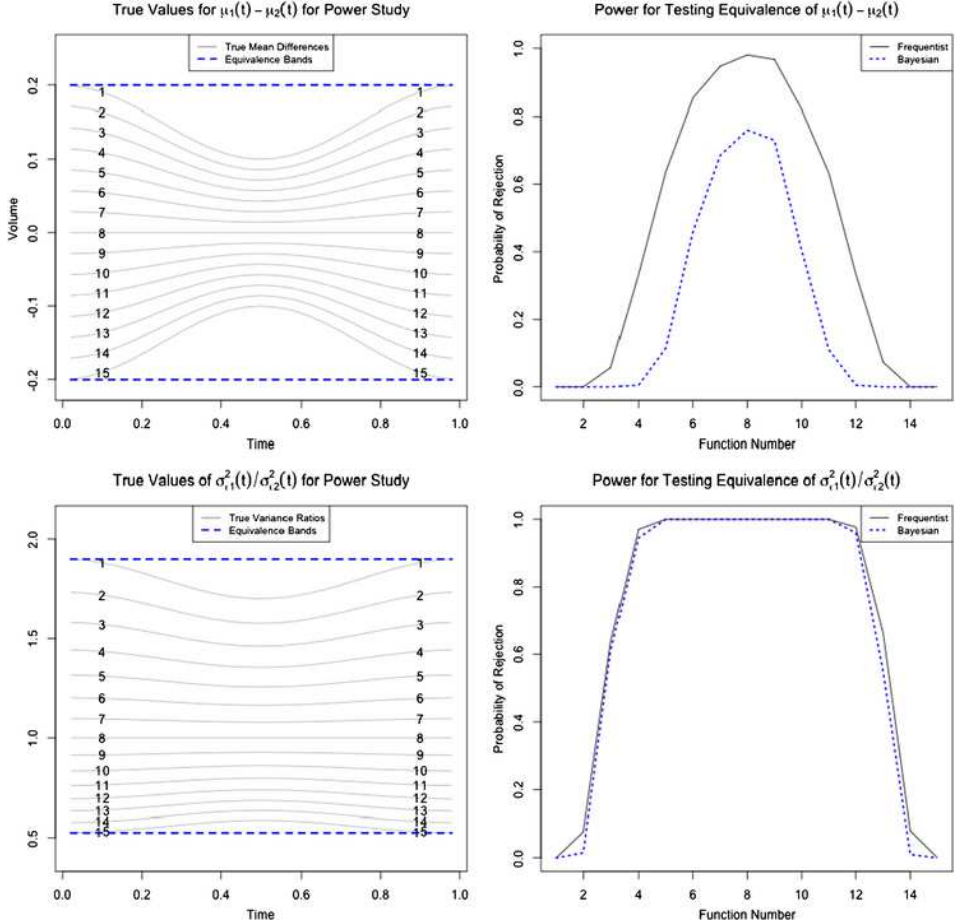}

\caption{Sequence of true values and corresponding power of $\mu
_1(\cdot
) - \mu_2(\cdot)$
(top) and $\sigma^2_{\epsilon,1}(\cdot)/\sigma^2_{\epsilon
,2}(\cdot)$
(bottom) along with upper and lower equivalence bands used for the
power study.}\label{figpowersequence}
\end{figure}

To investigate the power of our methods, we define a sequence of true
values for our metrics of equivalence that fall entirely between the
upper and lower equivalence thresholds. These sequences and numerical
labels are shown in Figure~\ref{figpowersequence}. The rest of our
simulation procedure mirrors that of our simulation for testing the
Type I error rate. Figure~\ref{figpowersequence} shows the results of
this study. We see that for testing equivalence of means, the
frequentist procedure appears to be substantially more powerful than
its Bayesian counterpart. For testing equivalence of variances, the
frequentist and Bayesian procedures behave quite similarly, with no
clear indication that one procedure is any more powerful than the other.
%
%

\section{Discussion}
We have presented a broad framework for equivalence testing when one's
data are intrinsically functional. This framework begins with
definitions of metrics of equivalence, and correspondingly with the
establishment of upper and lower equivalence bands which are themselves
functions of the continuum over which the functional data is defined.
We have stressed the importance of using metrics that are able to
discern similarity of location and of spread, as neither individually
is sufficient for suggesting equivalence. We illustrated the proper use
of these frameworks using data from a method comparison study assessing
the performance of a new device for testing pulmonary function, SLP,
relative to the gold standard for pulmonary diagnoses, the spirometer.

Our model presently makes an assumption that all individuals are drawn
from the same population. For our application this makes sense, as we
are solely looking at healthy individuals. For other applications, the
individuals for which repeated measurements are attained may be draws
from multiple populations. In our application, one could potentially
have individuals of varying degrees of pulmonary health (e.g., healthy,
asthmatic, smokers). Our model can easily adapt to this, as this simply
requires adding an additional level to the hierarchy. We could either
say that health level specific means are drawn from a population with
an overall mean, and then individual means are drawn from these health
level specific populations, or we could model the health level means as
fixed effects and result in a functional mixed effects model.

Using the difference between mean functions to test locational
disparity is a natural choice, and the extent to which magnitude of
differences are important can be controlled by tightening or loosening
the equivalence bands. For testing the disparity between variances of
both errors and random effects, we have followed the prevalent choice
in the scalar equivalence testing literature [see \citet{liu92}] and
have used the ratio between variances, $\sigma^2_{1}(\cdot)/\sigma
^2_{2}(\cdot)$. On the one hand, this unitless measure has appeal in
that it has potential for standardization across applications. On the
other hand, we lose a sense of the absolute difference between the
quantities. For some applications, the difference between a variance of
0.01 and 0.02 could be inconsequential, yet the difference between 0.04
and 0.08 could be enough to warrant using one device over another. If
one were using ratios for assessing a discrepancy, however, these
quantities would be identically different. We thus suggest that the
difference between variances, $\sigma^2_1(\cdot) - \sigma^2_1(\cdot)$,
may be an additional metric for equivalence that could be used in
tandem with the ratio of variances to test for equivalence of variability.

Note that there may be additional facets of the underlying
distributions of functions to be addressed beyond location and
variability, depending on the application. For example, one may be
interested not only in the difference in the mean functions being
within an equivalence region, but also in the derivative of the
difference between mean functions being small in absolute value. We
leave the development of proper methodology for these questions as a
topic for future research, but the strategy of supplying upper and
lower equivalence bands would certainly be appropriate.

We hope that this paper serves as a valuable contribution to the
literature on equivalence testing and that its extension to the realm
of functional data will be useful for a host of applied users,
including but not limited to practitioners looking to compare devices
whose measurements cannot be summarized as scalar quantities.
Comparison studies are of the utmost importance, as oftentimes the
emergence of newer and better devices can have salubrious outcomes for
society in general. Our goal is that this paper properly emphasizes the
importance of equivalence testing in general, and provides traction for
researchers who aim to suggest that two populations of functions are
practically equivalent rather than to suggest that they are different.

\section*{Acknowledgments}
We would like to thank the anonymous Associate Editor and referees for
their invaluable feedback during the review process, which led to
marked improvements in the quality of the article. We would also like
to thank Dr.~Joan Lasenby and Stuart Bennet of the Signal Processing
and Communications Laboratory, Department of Engineering, University of
Cambridge for introducing us to the problem and providing us with the
data set used herein.


\begin{supplement}
\stitle{Supplement to ``Equivalence testing for functional data with an application to
comparing pulmonary function~devices''\\}
\slink[doi]{10.1214/14-AOAS763SUPP} 
\sdatatype{.pdf}
\sfilename{aoas763\_supp.pdf}
\sdescription{We provide a description of the preprocessing that our
data underwent, a~detailed derivation of our\break Metropolis-within-Gibbs
sampling algorithm, and diagnostic plots showing convergence of our
Gibbs sampler when used on our data.}
\end{supplement}

%

\printaddresses

\end{document}